\documentclass[a4paper,fleqn]{cas-sc}

\usepackage[numbers]{natbib}

\def\tsc#1{\csdef{#1}{\textsc{\lowercase{#1}}\xspace}}
\tsc{WGM}
\tsc{QE}

\begin{document}
\let\WriteBookmarks\relax
\def\floatpagepagefraction{1}
\def\textpagefraction{.001}

\shorttitle{<Effects of Cl and Zn co-doping on the performance of $\gamma$-CuI scintillation detection>}    

\shortauthors{<Chao Li, Meicong Li, Zhuli Zhang, Qiang Zhao, Naixin Liu,  Kailei Wang, Fan Zhang, Xiaoping Ouyang>}  

\title [mode = title]{ Chlorine and zinc co-doping effects on the electronic structure and optical properties of $\gamma$-CuI}  



%

\author[1]{Chao Li}[style=chinese]





\credit{Investigation, Methodology, Writing - original draft}

\affiliation[1]{organization={Department of Physics, Changzhi University},
            city={Changzhi},
            postcode={046011},
            country={China}} 
            
\author[2,3]{Meicong Li}[style=chinese]





\credit{Investigation, Methodology, Writing - original draft}

\affiliation[2]{organization={Beijing Key Laboratory of Passive Safety Technology for Nuclear Energy, North China Electric Power University},
            city={Beijing},
            postcode={102206},
            country={China}}
            
\author[1]{Zhuli Zhang}[style=chinese]





\credit{Methodology, Validation}


\author[2,3]{Qiang Zhao}[orcid=0000-0003-4719-6808]

\cormark[1]


\ead{qzhao@ncepu.edu.cn}


\credit{Conceptualization, Writing - review \& editing}

\affiliation[3]{organization={School of Nuclear Science and Engineering, North China Electric Power University},
            city={Beijing},
            postcode={102206},
            country={China}}

\cortext[1]{Corresponding author}


\author[4]{Naixin Liu}[style=chinese]





\credit{Visualization, Validation}

\affiliation[4]{organization={Advanced Ultraviolet Optoelectronics Co., Ltd.},
            city={Changzhi},
            postcode={046000},
            country={China}}

\author[1]{Kailei Wang}[style=chinese]




\credit{Formal analysis, Validation}


\author[1]{Fan Zhang}[style=chinese]

\cormark[2]

\ead{zhangfan@mail.bnu.edu.cn}


\credit{Formal analysis, Funding acquisition}

\cortext[2]{Corresponding author}

\author[5]{Xiaoping Ouyang}[style=chinese]




\credit{Supervision}

\affiliation[5]{organization={Northwest Institute of Nuclear Technology},
            city={Xi'an},
            postcode={710024},
            country={China}}

\begin{abstract}
The effects of chlorine (Cl) and zinc (Zn) co-doping on the electronic structure and optical properties of the zinc blende ($\gamma$) phase of copper iodide ($\gamma$-CuI) scintillator material are investigated by using first-principles density functional theory calculations. The band structure, density of states, dielectric function, absorption coefficients, and reflectivity were analyzed before and after doping. Results show co-doping significantly modifies the band structure, reduces the band gap, and generates impurity energy levels. Cl doping enhances absorption in the high energy region while reducing visible light absorption. Zn doping induces a redshift in absorption and n-type conductivity at high concentrations. With suitable co-doping ratios, the absorption coefficient and reflectivity of $\gamma$-CuI can be optimized in the visible range to improve scintillation light yield. The calculations provide guidance for co-doping $\gamma$-CuI scintillators to achieve superior detection performance. The n-type conductivity also makes doped $\gamma$-CuI promising for optoelectronic applications. 
\end{abstract}


\begin{highlights}
\item The study investigates the effects of chlorine and zinc co-doping on the electronic structure and optical properties of $\gamma$-CuI.
\item Co-doping significantly modifies the band structure and improves the optical absorption properties of $\gamma$-CuI.
\item Co-doped $\gamma$-CuI has potential applications in optoelectronic devices and solar cells.
\end{highlights}


\maketitle

\section{Introduction}\label{}



Cuprous iodide (CuI) is widely used in different applications such as life science reagents industry due to its antifungal and bactericidal properties, food \& feed additive industries to improve animal growth and health, pharmaceuticals, photography, and pigments. It is also commonly used as an ionic conductive material in solid-state batteries, high-performance solar cells, and thermoelectric devices. CuI can also be used as a scintillator material, which has the following advantages\cite{lin2016luminescence,he2023copper}: 1) CuI crystal has high luminescence efficiency and energy resolution; 2) CuI crystal has short luminescence time and fast response speed; 3) The preparation process of CuI crystal is simple and low cost; 4) CuI crystal has low toxicity and no pollution to the environment.

Since 1907, when Bädeker iodized metallic copper by iodizing it in iodine vapor, and reported the transparent semiconductor properties of $\gamma$-CuI firstly \cite{baedeker1906elektrische}, and then the research work on $\gamma$-CuI has been carried out. CuI has a rich phase diagram, meaning that it exists in several crystalline forms. It adopts a zinc blende structure below 390 °C ($\gamma$-CuI), a wurtzite structure between 390 and 440 °C ($\beta$-CuI), and a rock salt structure above 440 °C ($\alpha$-CuI) .  Grauinyt et al. reported that $\gamma$-CuI has a relatively large band gap (3.1 MeV), a relatively high transmittance and hole electrical conductivity\cite{grundmann2013cuprous}. Yue et al. \cite {yue2017effect} showed that, so far found, $\gamma$-CuI is the fastest inorganic scintillator, the decay time is fast only 130 ps, so it has a wide application in ultrafast radiation measurement, and it found that the addition of I and Zn can significantly enhance the luminescence at 435 nm and depress the luminescence near 680 nm. Zhang et al. \cite {Zhang2019Cl}conducted an experiment that showed the oxidation problems of the chlorine-doped growth liquid and the crystals are greatly improved by the introduction of an argon atmosphere. Additionally, the ultrafast near-band edge emission of $\gamma$-CuI:Cl crystals is significantly enhanced. The current research on $\gamma$-CuI is mainly focused on single element doping. Hao et al. \cite {hao2021tuning} used anion doping to regulate the intrinsic defects in $\gamma$-CuI.  Graužinyt\.{e} et al. \cite {grauvzinyte2019computational} explored the possibility of enhancing the performance of $\gamma$-CuI by impurity doping using a high throughput method. Chen et al. \cite{chen2019defect} studied the structure and electronic properties of group IIB doped CuI by first  principles calculation. Zhu et al. \cite { GU2013} studied the structure and electronic properties of Zn, Ga and Al doped CuI. These studies showed that element doping can change the properties of CuI, but all these research works focused on single elements doping, and the effects of elemental co-doping on $\gamma$-CuI has rarely been reported.

The doping process involves intentionally introducing impurities into a material to alter its electrical conductivity, carrier concentration, and other properties\cite{muhammad2018ab,kalwar2022geometric}. Co-doping takes this process one step further by introducing two or more types of impurities into the material simultaneously. The interaction between these dopants can result in synergistic effects that lead to enhanced properties beyond what would be achieved by doping with a single impurity\cite{rafique2019theoretical}. co-doping of semiconductors with donor and acceptor impurities can create a p-n junction, which is the basis of many electronic devices such as diodes and transistors. Co-doping can also be used to modify the optical properties of materials, such as the bandgap, which is important for applications in solar cells and photodetectors. Khodyuk et al. showed enhancement of energy resolution and light output by Eu and Ga co-doping NaI:Tl\cite{khodyuk2015optimization}. Ito et al. reported that the non-proportionality in the low energy region was improved by Na and K co-doping, in the light yield proportionality under $\gamma$-ray irradiation\cite{ito2016effects}. Liu et al. found Mg and Ce co-doping Lu$_3$Al$_5$O$_{12}$ leads to a significant decrease of thermoluminescence intensity above room temperature and an increase of scintillation light yield (LY) value and fast component content even if the overall scintillation efficiency decreases\cite{liu2014effect}.  Alekhin et al. reported that Sr and Ca co-doping can improve the $\gamma$-ray energy resolution of LaBr$_3$:Ce\cite{alekhin2013improvement}. Demircan et al. studied the effect of Co and Mn Co-doping on structural and optical propertied of ZnO thin films. They found Mn(II) and Co(II) metals are bonded with oxygen atoms in the ZnO structure, and this proves that the energy holding capacity is increased due to the red region shift\cite{demircan2022effect}. Kamada et al. found both Mg-Ce and Ca-Ce co-doping Gd$_3$Al$_2$Ga$_3$O$_{12}$ can accelerate the scintillation decay\cite{kamada2015alkali}. Nishimoto et al. found that the light yield of the La co-doped Eu7.5\%:SrI$_2$ specimens were lower than that of the Eu7.5\%:SrI$_2$ specimen due to the generation of some defects by La co-doping\cite{nishimoto2014effects}. 

Computer simulation methods play a very important role in scientific research and can provide important theoretical support and experimental guidance for many fields such as materials science, physics, and chemistry. With the development of computing power, the first-principles calculation method has been widely applied in scientific research, especially in the research of novel scintillator materials. Using a first-principles approach, several elements have been found that can be used as activators in scintillators by doping\cite{bang2013suppression, mcallister2015auger, wu2016first, schleife2017excitons}. Specifically for halides, current research on co-doping has focused on the use of halide perovskite as scintillator materials\cite{yao2020enhanced, deng2021electronic, bhamu2021improving, sampson2017transition, chen2018all, zhang2021reproducible}. We believe that the first-principles calculations can also be used to investigate the effect of co-doping on the use of CuI as a scintillator detector material.

In this work, we choose Cl and Zn elements as the doping elements. Based on our previous researches\cite{zhang2016electronic, zhao2017electronic, jiang2017electronic, zhang2021effects, li2023electronic}, we investigated the effects of Cl doping, Zn doping, and Cl and Zn co-doping (1:1; 1:2; 2:1, respectively) on the electronic structure and optical properties of $\gamma$-CuI using a first principles calculation method. We calculated the lattice constants, formation energy, dielectric function, optical absorption coefficients and dielectric functions before and after doping.

\begin{figure}[htpb]
	\centering
		\includegraphics[width=12.0cm]{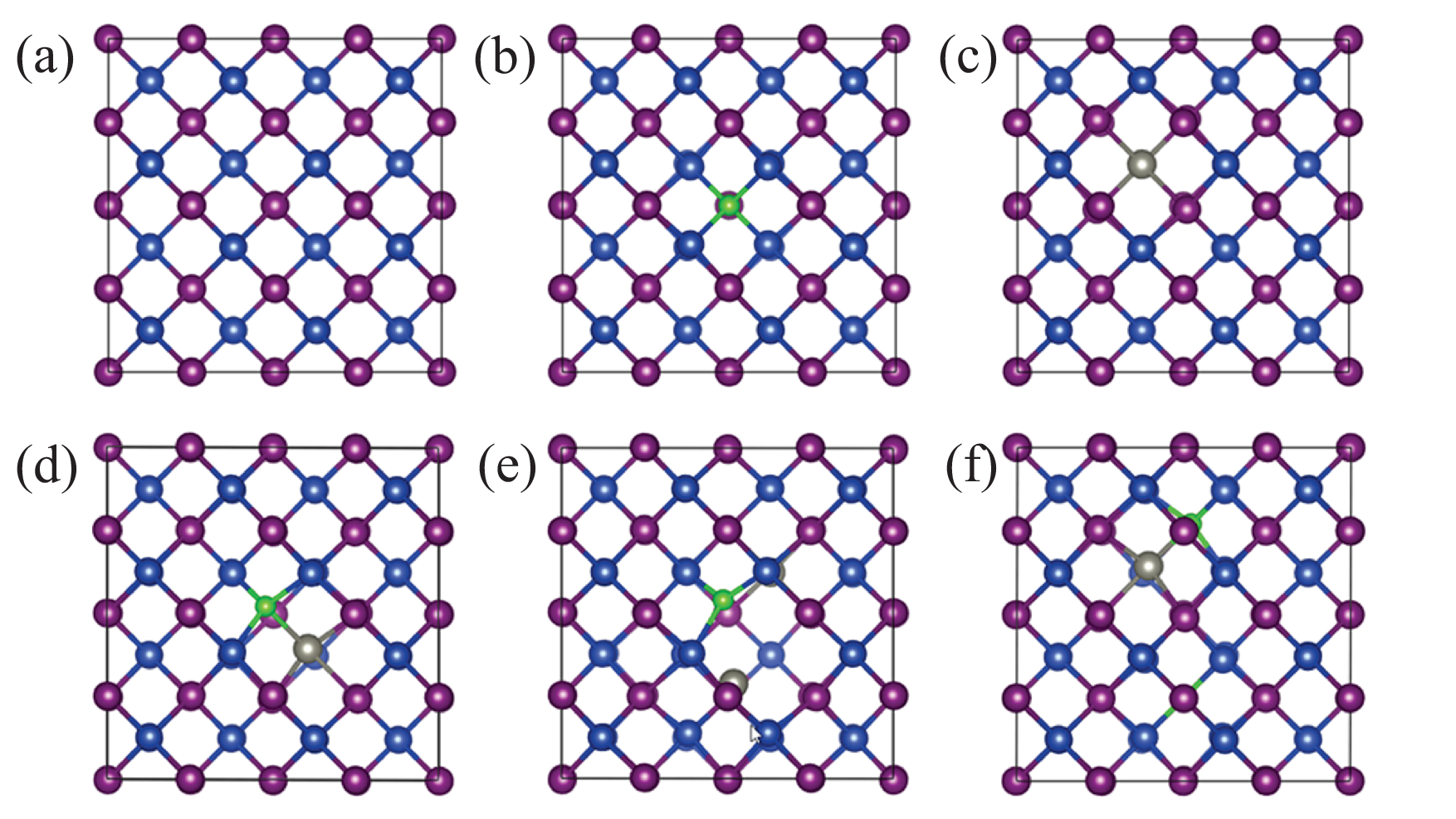}
	  \caption{Schematic representation of the doped $\gamma$-CuI structure, where each atom type is denoted by a different color: copper (Cu) atoms are shown in blue, iodine (I) atoms in purple,  and dopant atoms chlorine (Cl) in green, zinc (Zn) in grey.}\label{fig1}
\end{figure}




\section{Calculation method}\label{}

All the calculations in this paper were carried out with the Vienna Ab initio Simulation Package (VASP) code\cite{kresse1996efficient,kresse1996efficiency} based on the density functional theory (DFT). The Projector-Augmented Wave (PAW) pseudopotentials method was used to describe the interactions between atomic core and valence electrons. The exchange-correlation potential was represented by the Generalized Gradient Approximation (GGA) in the form of the Perdew-Burke-Ernzerhof (PBE) functional\cite{perdew1996generalized}. The electron wave function was expanded in a basis set of plane waves with a kinetic energy cutoff of 400 eV. The position of all the atoms was fully relaxed in the configuration optimization process, the Hellman–Feynman force was less than 0.02 eV/Å, and electronic iterations convergence was set to 1.0$\times$$10^{-5}$ eV, Monkhorst–Pack $k$-point meshes of  3 $\times$ 3 $\times$ 3 was used for both geometric optimization and self-consistent calculations. 

In order to study the effects of Cl and Zn doping and co-doping on the electronic structure and optical properties of $\gamma$-CuI, we build a 2 $\times$ 2 $\times$ 2 supercell structure as shown in Figure 1. Based on the results of previous research\cite{grauvzinyte2019computational}, substitutional doping is easily achieved on Cu-site and I-site. We replaced the I atom with a Cl atom and replaced the Cu atom with a Zn atom. In this work, we mainly consider the effects of doping with different concentrations of Cl and Zn elements and co-doping with different ratios of their on the properties of $\gamma$-CuI.  Therefore, during co-doping, it is still Cl atoms replacing I atoms and Zn atoms replacing Cu atoms. Their relative positions are determined by the Wigner-Seitz method\cite{wigner1933constitution}. We calculated parameters such as lattice constant, density of states, energy band structure and optical absorption coefficient in order to analyze the calculated results before and after doping. Figure 1 (a), (b), (c), (d), (e), and (f) denote intrinsic $\gamma$-CuI, Cl-doped, Zn-doped, Cl:Zn-doped, Cl:2Zn-doped, and 2Cl:Zn-doped, respectively. Cl:Zn stands for chlorine to zinc doping ratio of 1:1, Cl:2Zn stands for chlorine to zinc doping ratio of 1:2, 2Cl:Zn stands for chlorine to zinc doping ratio of 2:1, where blue stands for copper atom, purple stands for iodine atom, green stands for chlorine atom and gray stands for zinc atom.

\section{Results and Discussion}\label{}
\subsection{Lattice constants}

\begin{figure}[htpb]
	\centering
		\includegraphics[width=12.0cm]{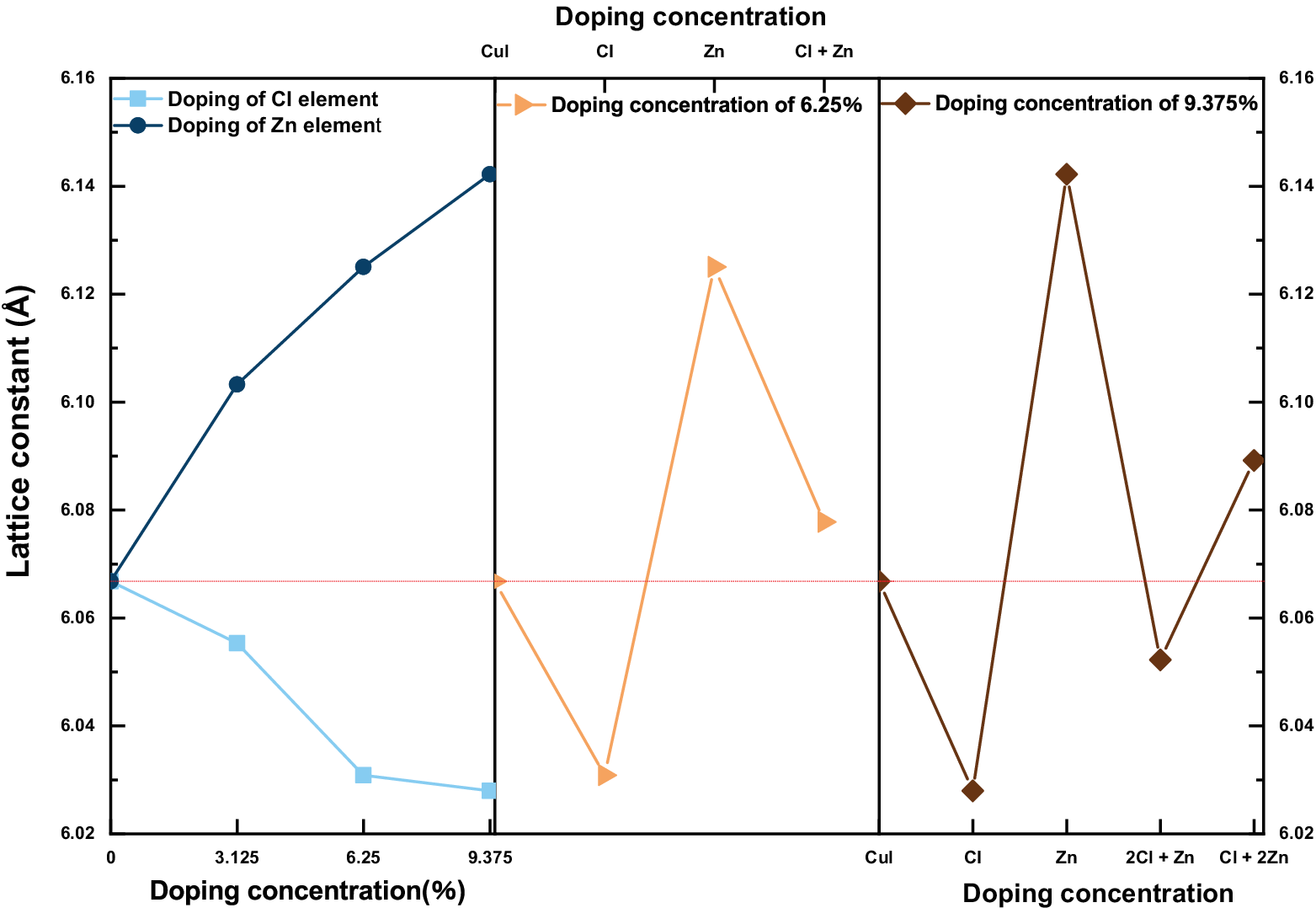}
	  \caption{Effect of elemental doping on the lattice constant of $\gamma$-CuI.}\label{fig2}
\end{figure}

The lattice constant is a parameter that reflects the basic structure of a crystal substance, and the change of in the lattice constant reflects the change in the composition and the state of force inside the crystal. Firstly, we calculated the lattice constant of the $\gamma$-CuI by the first principles calculation method, and the results are shown in Table 1.The lattice constant a of $\gamma$-CuI is 6.067 Å, which is in good agreement with the experimental results \cite {hull1994high,yashima2006crystal} as well as the theoretical data \cite {ma2004first, amrani2006first, chen2013first, yadav2014first}. This shows that our choice of computational details gives good results on structure parameters.

Then, we also simulated the $\gamma$-CuI lattice constants before and after doping, and the results are shown in Figure 2, from left to right three panels indicate the effect of doping with different concentrations of Cl and Zn on $\gamma$-CuI lattice constant, the effect of doping with equal proportions of Cl and Zn on $\gamma$-CuI lattice constant and the effect of doping with different proportions of Cl and Zn on $\gamma$-CuI lattice constant, respectively. It can be clearly seen that the doping of Cl element makes the lattice constant smaller than $\gamma$-CuI, and it decreases with increasing doping concentration. This is because the radius of Cl atoms is smaller I atoms, but the effect on the lattice constant becomes smaller when the doping concentration increases from 6.25 at\% to 9.375 at\%, which is probably due to the increasing doping concentration of Cl elements make the interaction between Cl atoms stronger and thus reduces the effect on the $\gamma$-CuI lattice constant to some extent. The Zn atom, however, has a larger atomic radius than the Cu atom, which makes the doping lattice constant larger and increases with increasing doping concentration. When the two elements of Cl and Zn are co-doped, equal proportional doping of Cl and Zn elements result in a larger lattice constant, it shows that the doping of Zn elements has a greater effect on the $\gamma$-CuI lattice constant. When the concentration of doped Cl atoms is larger than that of Zn atoms, the lattice constant decreases; when the concentration of doped Cl atoms is smaller than that of Zn atoms, the lattice constant increases. In summary, we think there are two reasons for the change of the lattice constant of $\gamma$-CuI: 1) the different atomic radii; 2) the change of the interatomic force state before and after doping. We can control the lattice constant of $\gamma$-CuI by the doping of Cl and Zn or changing the doping ratio of them, so that we can get the crystals with suitable size structure.
In order to investigate the effect of doping elements on the properties of $\gamma$-CuI scintillation detectors, the electronic structure and optical properties before and after doping have been calculated. The next, we will discuss the energy band structure, density of states, and light absorption coefficient.

\begin{table}
\centering
\caption{The lattice constant $a$ (in \AA) of the $\gamma$-CuI.}
\label{table1}
\begin{tabular}{cc}
\hline
    lattice constant   &$a$ \\
\hline
    This work          &6.067 \\
    Experimental data  &6.05\cite{hull1994high}, 6.058\cite{yashima2006crystal}\\ 
    Theoretical results&6.09\cite{ma2004first},6.097\cite{amrani2006first},6.073\cite{chen2013first},6.05\cite{yadav2014first}\\
\hline
\end{tabular}
\end{table}

\subsection{Electronic structure}

\begin{figure}[htpb]
	\centering
		\includegraphics[width=12.0cm]{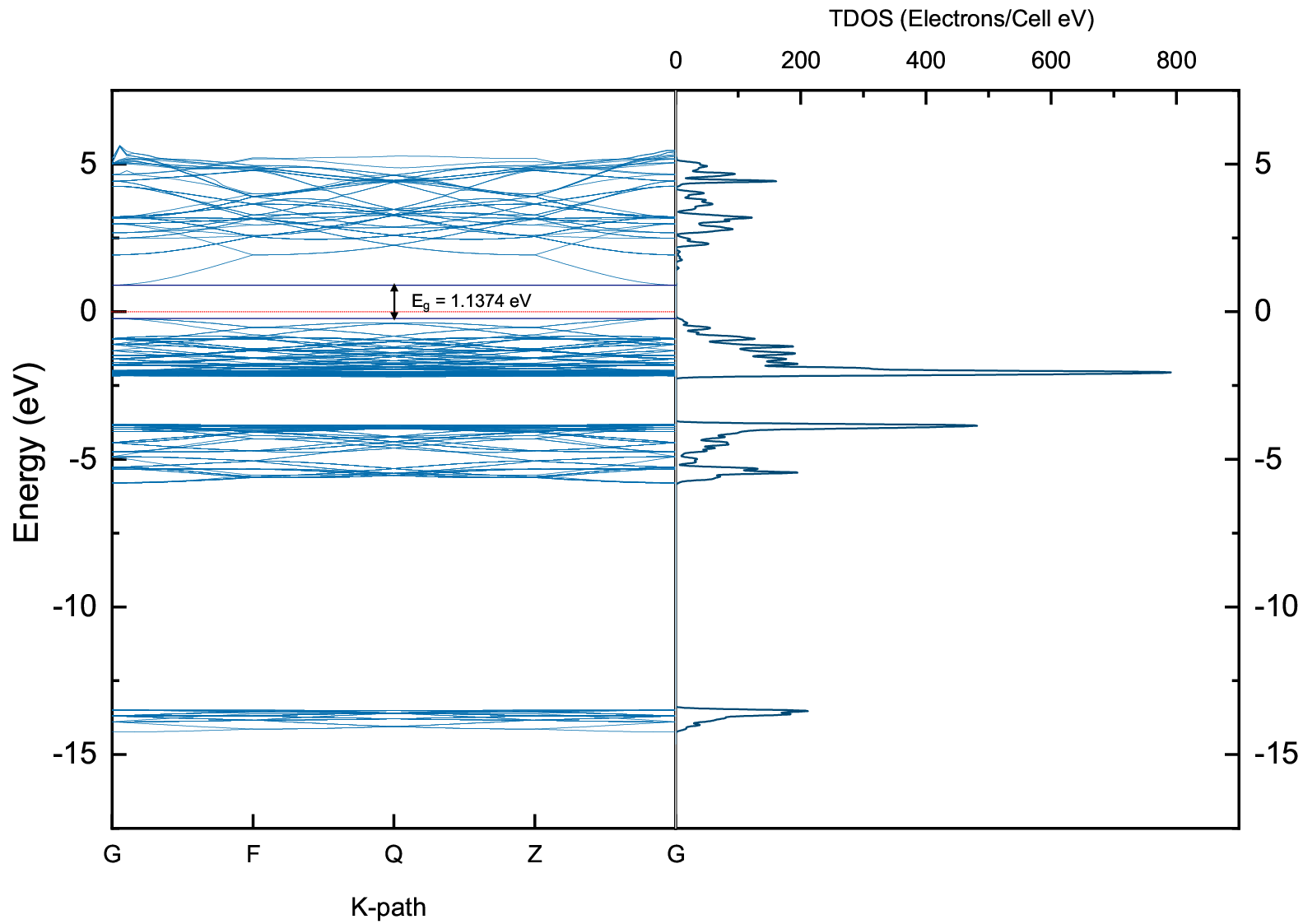}
	  \caption{The Band structure and DOS of $\gamma$-CuI.}\label{fig3}
\end{figure}

We calculated the intrinsic $\gamma$-CuI energy band structure and the density of states, as shown in Figure 3. It shows the band gap $E_g$ = 1.1374 eV for $\gamma$-CuI, and this value is in good agreement with previous theoretical results\cite{ma2004first,amrani2006first, chen2013first,yadav2014first}, as shown in Table 2. It should be noted that the band gap of the $\gamma$-CuI scintillator calculated by the GGA calculation is lower than the experimental value, because the exchange–correlation interaction between the d and f electrons is not sufficiently described in the GGA calculations. Density functional theory calculated by the ordinary plane wave Kohn-Sham equation often produces a smaller band gap than observed in experiments. DFT+U and hybridization density functional can adjust the U value and hybridization ratio, respectively, to obtain a band gap consistent with the experiment, but these methods require empirical evidence to set the parameters. The GW method, which is computationally intensive, typically results in a larger band gap than the experimental value when fully converged. In this work, we focus on the relative energy change of the band gap of $\gamma$-CuI scintillator caused by doping rather than its absolute energy, which is the reason why we choose the GGA generalization in the form of PBE. The density of states plot on the right side of Figure 3 shows that the valence band near the top of the intrinsic $\gamma$-CuI is mainly contributed by the Cu 3d and I 5p states, while the conduction band near the bottom is mainly controlled by I 5p, which is also in agreement with the theory.

\begin{table}
\centering
\caption{The calculated band gap $E_g$ (in eV) for the pure $\gamma$-CuI system compared with experimental data.}
\label{table2}
\begin{tabular}{cc}
\hline
    Band gap   &$E_g$ \\
\hline
    This work          &1.137 \\
    Expt. data  &3.118\cite{gogolin1989piezobirefringence}\\
    GGA  &1.05\cite{vettumperumal2019analysis},\quad 1.12\cite{zhang2019electronic},\quad 1.165 \cite{wang2010origins}\\
    GGA+U &1.86 (U=4.8)\cite{zhang2020electronic},\quad 1.89 (U=5.2) \cite{huang2012first},\quad 2.1 (U=6.0)\cite{pecoraro2022first}\\
    HSE &2.59\cite{chen2013first}, \quad2.57\cite{grauvzinyte2019computational}, \quad3.05\cite{zhang2020electronic}\\
    GW &1.79\cite{van2017automation}, \quad2.70\cite{pishtshev2017structure},\quad 3.29\cite{zhang2019electronic}\\
\hline
\end{tabular}
\end{table}

\begin{figure}[htpb]
	\centering
		\includegraphics[width=12.0cm] {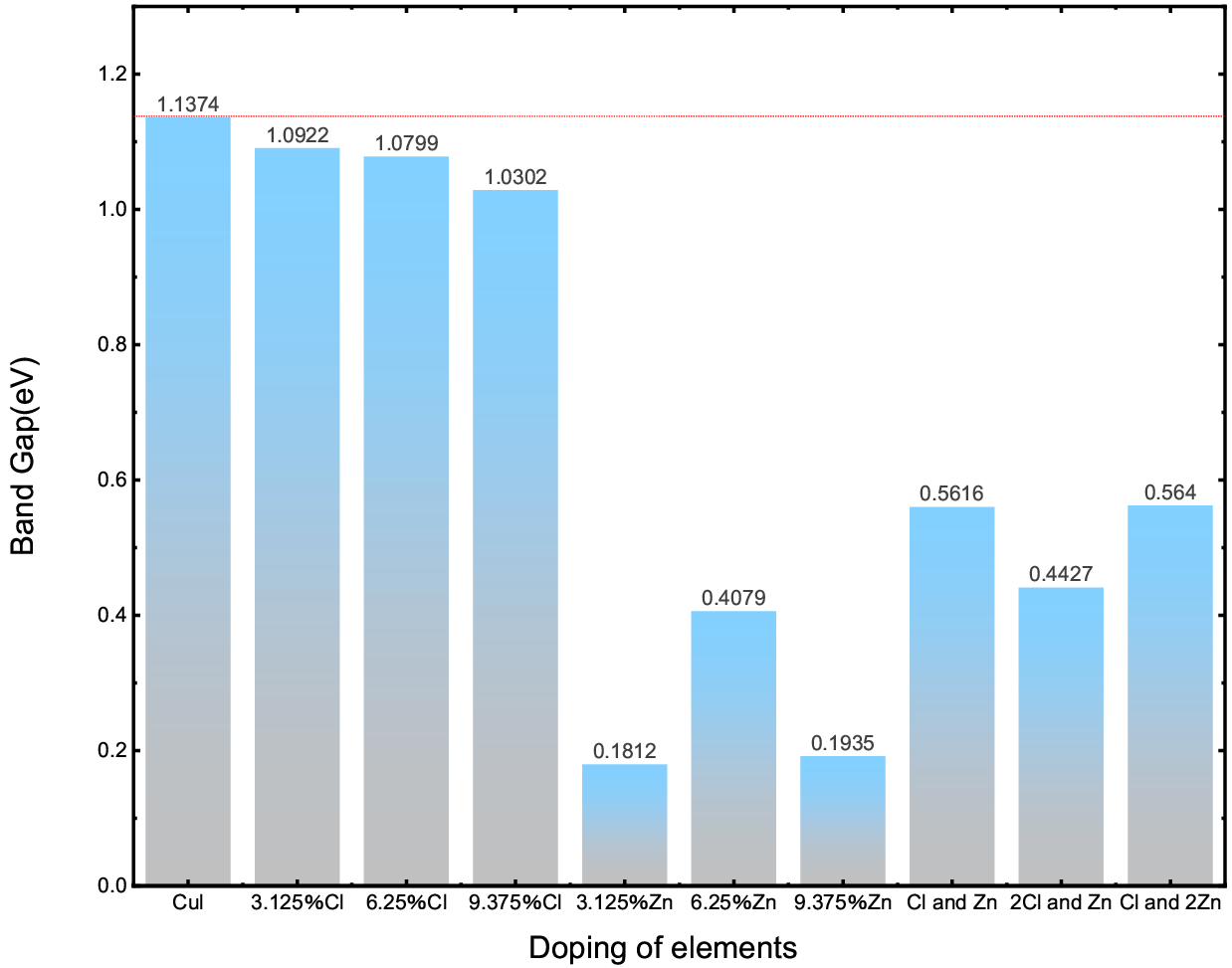}
	  \caption{The band gap of doping elements.}\label{fig4}
\end{figure}

In order to investigate the electronic structure properties of doping for $\gamma$-CuI, we analyze and discuss the energy band structure as well as the electronic density of states.
Figure 5 shows the structure of $\gamma$-CuI energy band before and after doping, where the red dashed line is the Fermi energy level. From left to right in Figure 5 (a), the energy band structures of the intrinsic $\gamma$-CuI, 3.123 at\% Zn element doping, 6.25 at\% Zn element doping and 9.375 at\% Zn element doping are shown. It can be seen that Zn element doping around -7.5 eV appears a new energy level, which is mainly contributed by the d state of the doped Zn element, so the number of energy levels here is increasing with the Zn doping concentration increasing. All the energy band structure moves down, the energy level density is higher so that the valence band energy level also increases.  Overall, the number of electrons that can occur energy level jump on the valence band is increasing, so that the probability of electron jumping is increasing, and improve the probability of being detected. In addition, the impurity energy level appears in the forbidden band, and with the increasing doping concentration of Zn elements, the Fermi energy level also enters from the valence band to the conduction band, thus the Fermi energy level is in the conduction band at 9.375 at\% Zn doping concentration, which behaves as an N-type semiconductor property. From Figure 4, it can be seen that the band gap is 0.1812 eV for 3.123 at\% Zn concentration, 0.4079 eV for 6.25 at\% Zn concentration and 0.1935 eV for 9.375 at\% Zn concentration, which increases and then decreases with the concentration increasing, but all are smaller than the band gap of the intrinsic $\gamma$-CuI.

Figure 5 (a) also shows a schematic diagram of the energy band structure of Cl element doping, from left to right, are 3.123 at\% Cl element doping, 6.25 at\% Cl element doping and 9.375 at\% Cl element doping, respectively. From Figure 4, the band gap are 1.0922 eV, 1.0799 eV, and 1.0302 eV in order, which means that the doping of Cl elements decreases the band gap and is anti-proportional to the doping concentration. In addition, the doped energy levels are also denser, the number of energy levels in the valence band is more, easier to jump, and then the smaller band gap can make it easier for electrons to jump, so after the Cl doping, the time required for the leap is shortened, which can improve the detection efficiency. Figure 5 (b) show the schematic diagram of the co-doping energy bands with different ratios of Cl and Zn elements. We can know, when Cl and Zn elements are co-doped, impurity energy levels appear in the forbidden band no matter what the doping ratio is. From Figure 4, it can be seen that when 3.123 at\% of Cl and 3.123 at\% of Zn are co-doped, the band gap is 0.5616 eV; When 3.123 at\% Cl and 6.25 at\% Zn are co-doped, the band gap is 0.564 eV; when 6.25 at\% Cl and 3.125 at\% Zn are co-doped, the band gap is 0.4427 eV. Thus it can be seen that when Cl and Zn elements are co-doped, increasing the doping concentration of Zn elements can increase the band gap and increasing the doping concentration of Cl can decrease the band gap.

In summary, the doping of Cl and Zn elements alone and together will reduce the band gap, and the band gap of Cl element doping is inversely proportional to the doping concentration; Zn doping causes the Fermi energy level to move from the valence band to the conduction band as the concentration increases.
Next, we analyze from the perspective of density of states.

\begin{figure}[htpb]
	\centering
		\includegraphics[width=16.0cm]{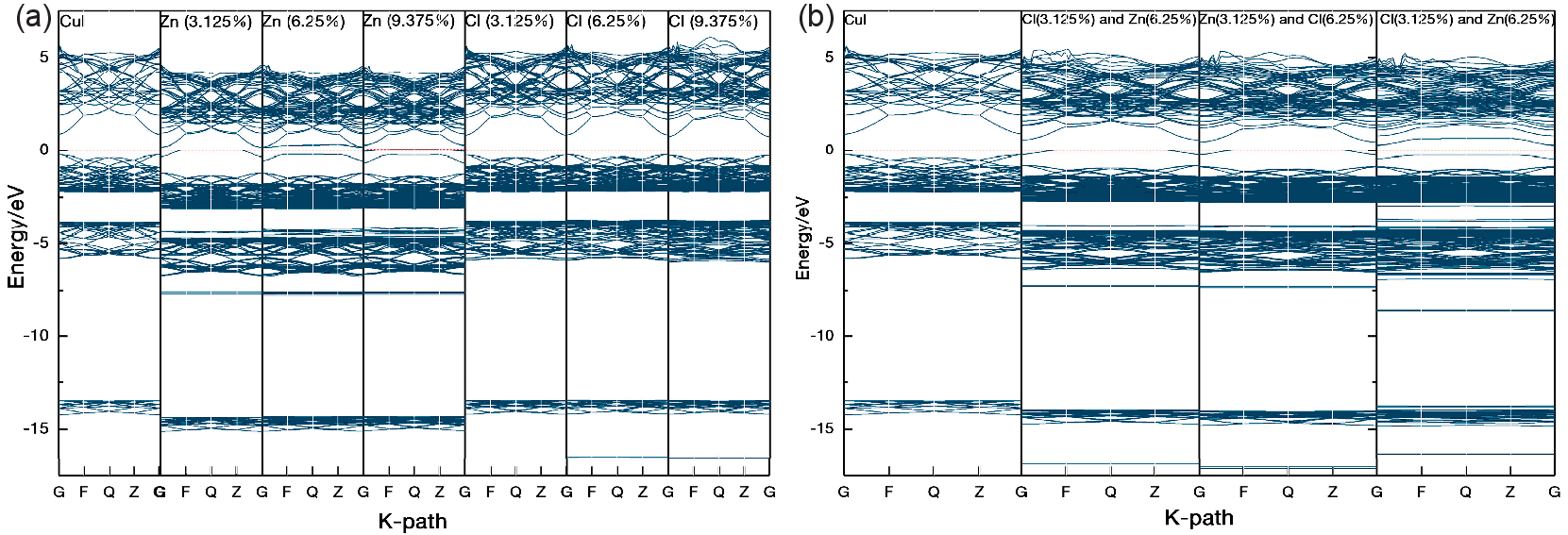}
	  \caption{The Band structure of $\gamma$-CuI and doping with Zn or Cl elements (a) and co-doping with Zn and Cl elements (b).}\label{fig5}
\end{figure}

\begin{figure}[htpb]
	\centering
		\includegraphics[width=16.0cm]{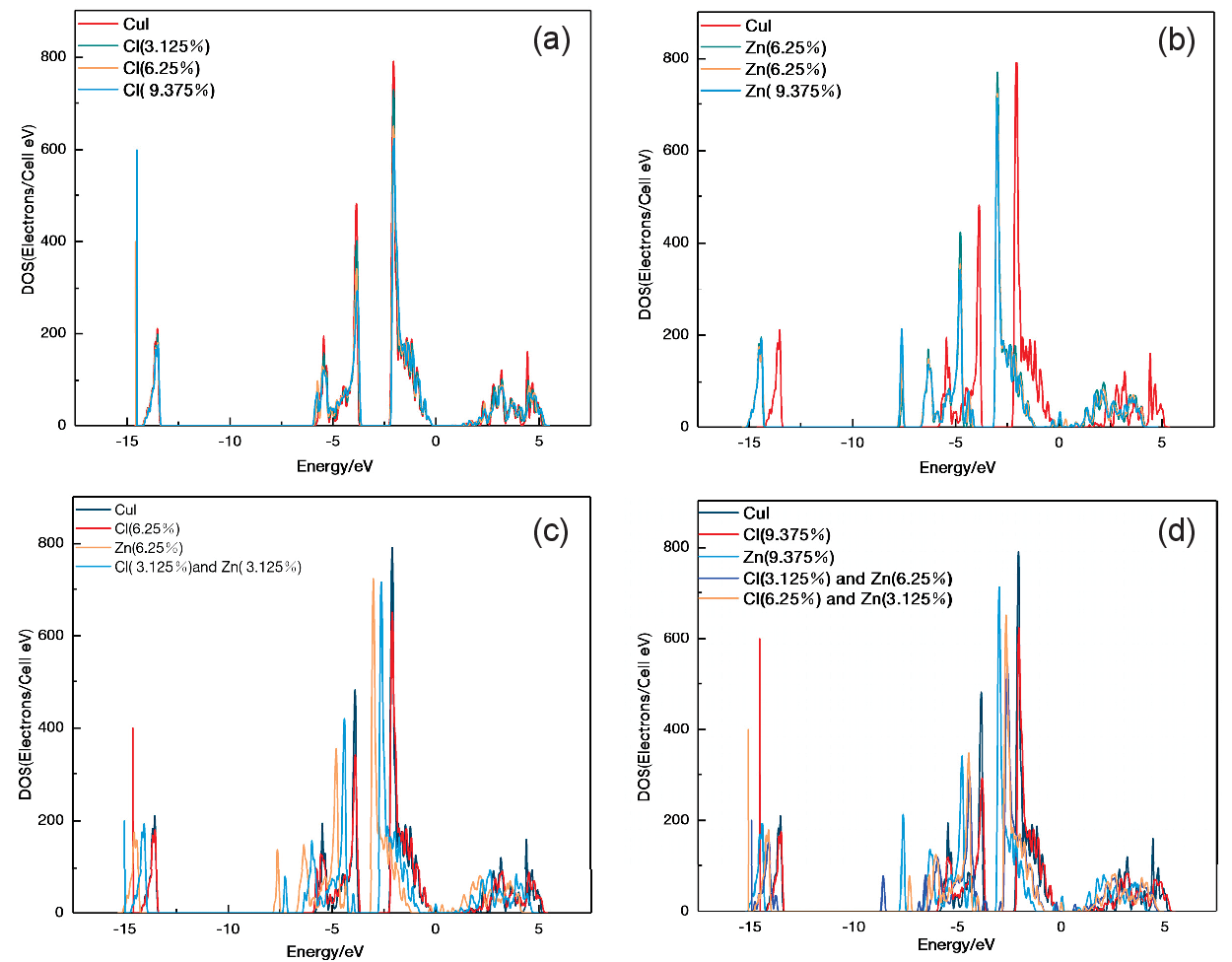}
	  \caption{The DOS of $\gamma$-CuI doping with Zn (a) or Cl (b) elements and co-doping of Zn and Cl elements (c)(d).}\label{fig6}
\end{figure}

\begin{figure}[htpb]
	\centering
		\includegraphics[width=12.0cm]{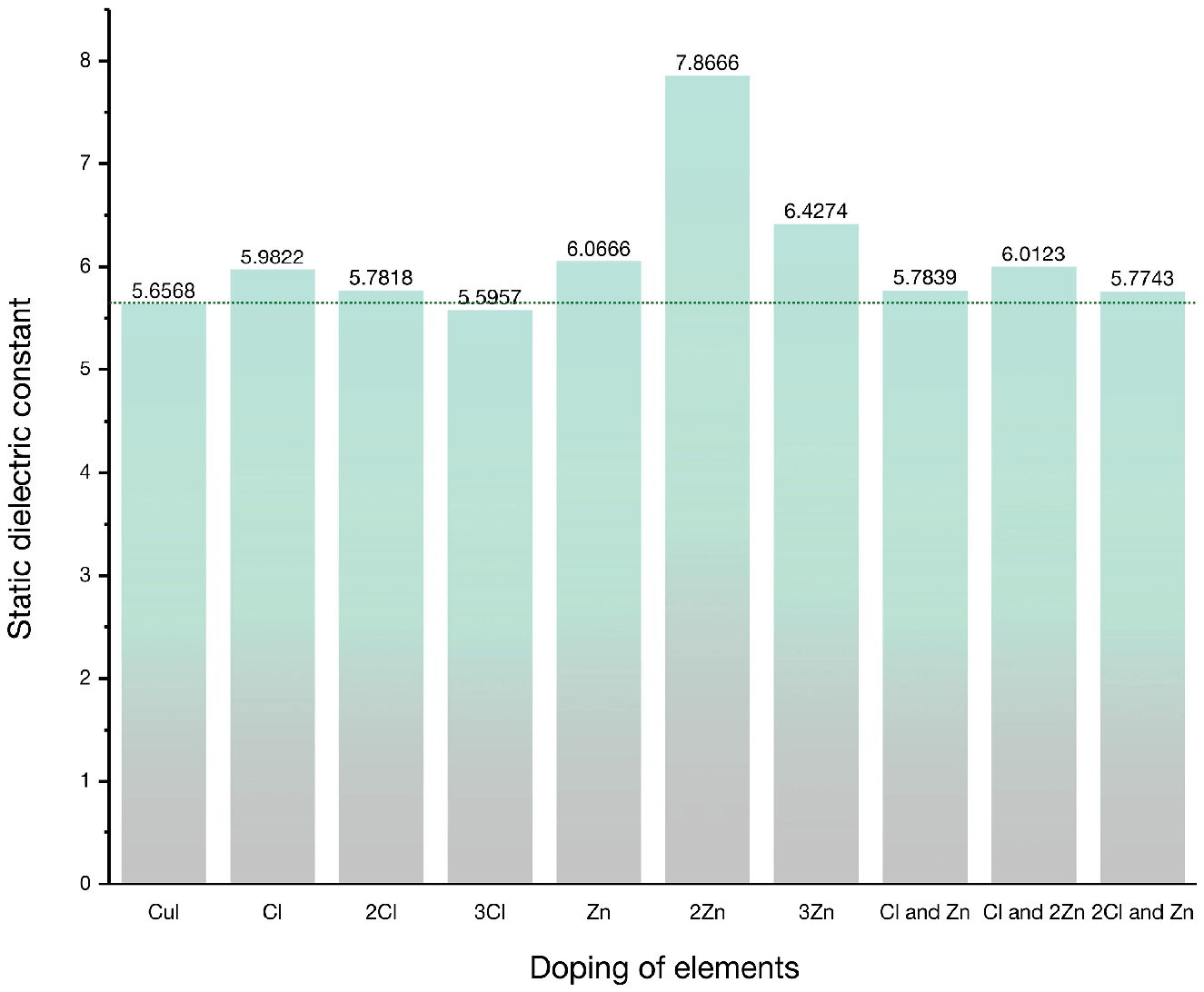}
	  \caption{The static dielectric constant of doping of elements.}\label{fig7}
\end{figure}

The density of states before and after doping are shown in Figure 6. It is very clear from Figure 6 (a) that the doping of Cl elements decreases the peak density of states, and is inversely proportional to the doping concentration. We find that the doping of Cl elements causes an abrupt change in the density of states around the energy of -14.6 eV, and unlike the density of states which decreases with increasing concentration, here the density of states increases with increasing doping concentration. This is mainly because the peak here is contributed by the s-state of Cl, so the variation is proportional to the doping concentration. Figure 6 (b) shows the effect of Zn doping on the density of states. It can be seen that the doping of Zn reduces the peak density of states, at the same time, shifts the density of states plot in the direction of energy reduction, that is, red shift occurs. However, as the doping concentration changes, the density of states does not shift, which means that the doping concentration is not related to the redshift. In besides, we find that similar to Cl doping, Zn doping also causes a new peak in the density of states, but near the -7.6 eV energy, and the peak here is contributed by the d state of Zn, so the variation is also proportional to the doping concentration. 

Next, we discuss the case of Cl and Zn co-doping, Figure 6 (c) and Figure show the schematic of density of states with different ratios of Cl and Zn co-doping. It can be seen from Figure 11 that the 1:1 ratio co-doping of Cl and Zn elements causes an overall move of the density of states in the direction of energy reduction, but the move is reduced relative to the doping of Zn elements alone, and it makes two new peaks of density of states appear, one at around -15 eV energy and the other at around 7.2 eV. Figure 6 (d) shows that changing the doping ratio between Cl and Zn elements, whether the Cl doping concentration is larger than the Zn doping concentration or the Cl doping concentration is smaller than the Zn doping concentration, the red-shift amount does not change relative to the equal doping of the them. However, when doped with 6.25 at\% Cl and 3.125 at\% Zn, the peak density of states is greater than 9.375 at\% Cl alone and less than 9.375 at\% Zn alone, and the density of states plot compared to 6.25 at\% Cl doping only a red shift occurs. That is, when Cl and Zn are co-doped 2:1, the Zn element doping mainly redshifts the density of states, and the effect of reducing the peak is weakened. When co-doped with 3.125 at\% Cl and 6.25 at\% Zn, the peak density of states is reduced both compared to the 9.375 at\% Zn element doping and to the 9.375 at\% Cl element doping.

In conclusion, we can know that the doping of Cl and Zn elements will reduce the peak density of states and is inversely proportional to the doping concentration; the doping of Zn elements will cause the density of states to move in the direction of energy reduction, but is not related to the doping concentration; the doping of Cl and Zn elements will cause the density of states to produce new peaks, but at different locations, which is in agreement with the conclusion that new energy levels appear in the energy band diagram, and the peak size is proportional to the doping concentration. When Cl and Zn are co-doped, the doping of Cl element will depress the red-shift produced by the doping of Zn element. When CCl>CZn, the doping of Zn element only plays the role of making the density of states red-shift; when $C_{\rm{Cl}}<C_{\rm{Zn}}$, the doping of Zn element enhances the reduction of the peak density of states by Cl element. ($C_{\rm{X}}$ is the concentration of doping element X.)

After discussing the energy band diagram and density of states diagram before and after doping, then we will discuss the optical properties before and after doping.

\subsection{Optical properties}
The relatively high luminosity, small light absorption coefficient and reflection coefficient are important parameters to evaluate whether it can be a good scintillation detector material or not. So in this paper, we have calculated the dielectric function, absorption coefficient and reflection coefficient before and after doping. 
The optical response function can be expressed as a complex dielectric function in the linear response range, and the equation is expressed as follows:

\begin{equation}\label{1}
\varepsilon(w)=\varepsilon_1(w)+i \varepsilon_2(w)
\end{equation}

where $\varepsilon_{1}(w)$is the real part of the dielectric function and $\varepsilon_2(w)$ is the imaginary part of the dielectric function. 
The real part of the dielectric function can be obtained from the Kramers-Kronig dispersion relation and the definition of the direct jump probability:

\begin{equation}\label{2}
\varepsilon_1=1+C_2 \sum_{C, V} \int_{B Z} \frac{2}{(2 \pi)} \frac{\left|e M_{C V}(K)\right|^2}{\left(E_C^K-E_V^K\right)} \frac{h^3}{\left(E_C^K-E_V^K\right)^2-h^2 \omega^2} d^3 K
\end{equation}

Dielectric function imaginary part:

\begin{equation}\label{3}
 \varepsilon_2=\frac{C_1}{\omega^2} \sum_{C, V} \int_{B Z} \frac{2}{(2 \pi)^3}\left|e M_{C V}(K)\right|^2 \delta\left[E_C^K-E_V^K-h \omega\right] d^3 K
\end{equation}
Absorption coefficient:
\begin{equation}\label{4}
I(\omega)=\sqrt{2} \omega\left[-\varepsilon_1(\omega)+\sqrt{\varepsilon_1(\omega)^2+\varepsilon_2(\omega)^2}\right]^{\frac{1}{2}}
\end{equation}

Where, $C_1=\frac{4 \pi^2}{m^2 \omega^2}, \quad C_2=\frac{8 \pi^2 e^2}{m^2}, \quad K(\omega)=\frac{1}{\sqrt{2}}\left\{\left[\varepsilon_1(\omega)^2+\varepsilon_2(\omega)^2\right]^{\frac{1}{2}}-\varepsilon_1(\omega)\right\}^{\frac{1}{2}}, \\ n(\omega)=\frac{1}{\sqrt{2}}\left\{\left[\varepsilon_1(\omega)^2+\varepsilon_2(\omega)^2\right]^{\frac{1}{2}}+\varepsilon_1(\omega)\right\}^{\frac{1}{2}}$.

$C$ denotes conduction band, $V$ denotes valence band; $e$ is the charge of the electron, $m$ is the electron mass, BZ denotes in the first Brillouin zone, $K$ is the inverse lattice vector, and $h$ is Planck's constant. $\left|e M_{C V}(K)\right|^2$ is the momentum leap matrix element, $\omega$ is the angular frequency, and $C_1$ and $C_2$ are constants. $E_C^K$ is the conduction band intrinsic energy level, $E_V^K$ is the valence band intrinsic energy level. 
Therefore, the absorption coefficient is strongly related to the dielectric function. Based on this, we have calculated the optical parameters before and after doping by first principles calculation, and next, we discuss and analyze the optical property parameters.
\subsubsection{Dielectric function}

Firstly, we perform data processing of the calculated results by vaspkit to obtain the imaginary and real parts of the dielectric function, where the real part of the dielectric function reflects the material's polarization ability\cite{kim1992modeling} and the imaginary part of the dielectric function characterizes the strength of the absorption of light by the medium. 

\begin{figure}[htpb]
	\centering
		\includegraphics[width=16.0cm]{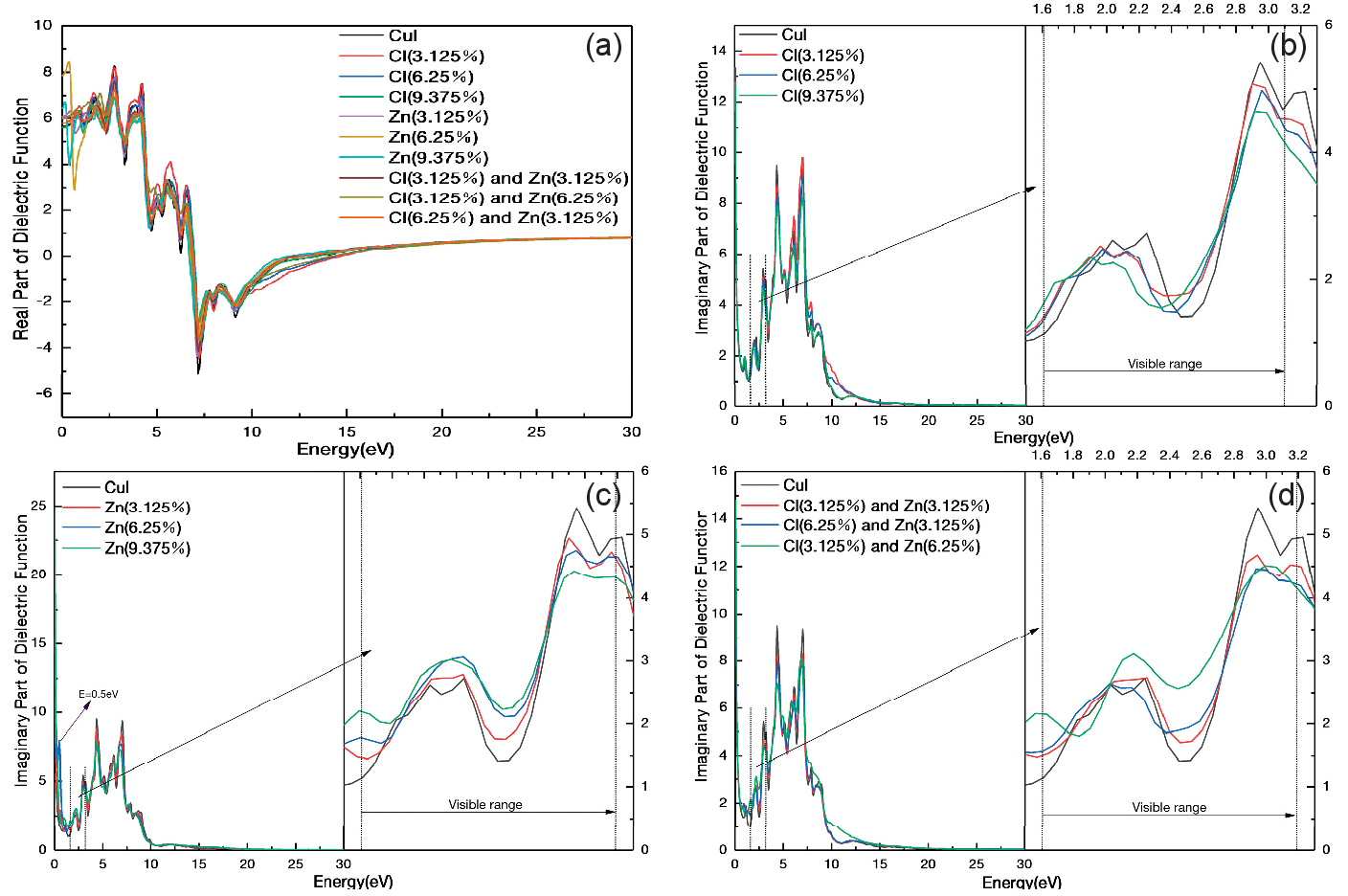}
	  \caption{The real and imaginary part of dielectriction function. Real part (a) and imaginary part for  Cl doping  (b), Zn doping (c) Cl and Zn co-doping (d)}\label{fig8}
\end{figure}

The static permittivity before and after doping is shown in Figure 7. It can be seen that, except for after the concentration of 9.375 at\% Cl element doping  the static dielectric constant is smaller than the intrinsic copper iodide, all other doping compared with the intrinsic copper iodide static dielectric constant is larger, and the concentration of 6.25 at\% Zn doping is the largest static dielectric function, then it has the strongest charge binding ability, the largest polarization ability, and the system of photogenerated electric field strength becomes larger, so that the photoexcited carriers in the crystal migration rate becomes faster, which shortens the detection time and then can improve the detection efficiency. The real part of the dielectric function before and after doping is shown in Figure 8 (a). The value of the real vertical coordinate of the dielectric function at energy 0 eV is the static dielectric function, and its magnitude reflects the polarizability of the material.

Figure 8 (b)-(d) show the imaginary part of our calculated dielectric function, with the imaginary part of the dielectric function in the visible range 1.61 to 3.19 eV on the right. The imaginary part of the dielectric function has a great relationship with the absorption of photons. The larger the imaginary part of the dielectric function is, the stronger the light absorption ability is. The left of Figure 8 (b) shows the imaginary part of the dielectric function that calculated from the doping of Cl elements. It can be seen that the imaginary part of the 3.125 at\% Cl concentration doping starts to be larger than the intrinsic $\gamma$-CuI when the energy is at larger than about 4.8 eV, while the 6.25 at\% and 9.375 at\% concentration doping starts to be larger than intrinsic $\gamma$-CuI when the energy is about larger than 7.2 eV. And in this energy range, the imaginary part of the dielectric function is decreasing with increasing doping concentration. In the visible range, as shown on the right, at energies less than 2.7 eV, the peak of the imaginary part of the dielectric function moves to the lower energy region, thus which makes the imaginary part in 1.61 eV to 1.97 eV is greater than the intrinsic $\gamma$-CuI, in 1.97 eV to 2.4 eV is less than the intrinsic $\gamma$-CuI, in 2.4 eV to 2.7 eV the imaginary part is greater than the intrinsic $\gamma$-CuI, and in 2.7 eV to 3.19 eV the imaginary part of the dielectric function is smaller than the intrinsic $\gamma$-CuI. In a words, the doping of Cl element makes the imaginary part of the dielectric function larger in the high-energy region, and the imaginary part of the dielectric function is reduced in the visible range, so that the absorption of the doped crystal is enhanced in the high-energy region for photons, the absorption coefficient is weakened in the visible range, which allows more photons within visible light to be detected. From this view, it can improve detection efficiency.

Figure 8 (c) shows the imaginary part of the dielectric function for Zn element doping. It can be seen that a new peak in the imaginary part of the dielectric function appears around 0.5 eV energy for 6.25 at\% Zn element doping, which means that the crystal doped with 6.25 at\% Zn element has a strong absorption here. The imaginary part of the dielectric function is smaller than the intrinsic $\gamma$-CuI at energies greater than 2.8 eV and is inversely proportional to the doping concentration. The right panel shows that in the visible range, the doping of Zn elements causes the imaginary part of the dielectric function to be larger in the energy range of 1.61 eV to 2.80 eV and increases with increasing concentration. In 2.80 eV to 3.19 eV, the imaginary part of the dielectric function is decreasing and is inversely proportional to the doping concentration. 

Figure 8 (d) shows the imaginary part of the dielectric function of the system after the co-doping of Cl and Zn elements. It can be seen that, when the energy is more than 2.7 eV, the imaginary part of the dielectric function is smaller for 3.125 at\% Cl and 3.125 at\% Zn elements co-doping compared to 6.25 at\% Cl elements doping and which is smaller than the intrinsic copper iodide, however, the dielectric function is larger when the energy is less then 2.7 eV. In the visible range, the imaginary part of the Cl and Zn co-doping is smaller than that of the intrinsic $\gamma$-CuI, which is not much different from that of Cl alone. When the proportion of Cl in the co-doping is increased, that is, when 6.25 at\% Cl and 3.125 at\% Zn elements are co-doped, the doping of Zn elements increases the imaginary part of the dielectric function, but it is still smaller than the imaginary part of the intrinsic $\gamma$-CuI in the visible range. While increasing the co-doping ratio of Zn elements, it increases the imaginary part of the dielectric function in the visible range and is larger than the imaginary part of the intrinsic $\gamma$-CuI in this range, but the peak at about 0.5 eV disappears due to the doping of the Cl element compared to the 6.25 at\% Zn concentration doping.

In conclusion, after Cl element doping, the imaginary part of the dielectric function is larger than the intrinsic $\gamma$-CuI in the high energy region (>4.8 eV) and smaller than the intrinsic $\gamma$-CuI in the visible range, Zn element doping at energies greater than 2.8 eV, where the imaginary part of the dielectric function is smaller than the intrinsic $\gamma$-CuI and inversely proportional to the doping concentration, while in the visible range makes the imaginary part of the dielectric function larger than the intrinsic $\gamma$-CuI. When Cl and Zn elements are co-doped, the imaginary part of the dielectric function in the visible range is larger than the intrinsic $\gamma$-CuI when the Zn concentration is larger than the Cl concentration. When the Zn concentration is smaller than or equal to the Cl concentration, the imaginary part of the dielectric function in the visible range is smaller than the intrinsic $\gamma$-CuI. Overall, when Cl elements are doped and CI and Zn elements are co-doped (the proportion of Zn elements doped is smaller than the proportion of Cl elements), the dielectric function is reduced, which can reduce the absorption in the visible range and improve the probability of being detected, thus improving the detection performance of $\gamma$-CuI.

\begin{figure}[htpb]
	\centering
		\includegraphics[width=12.0cm]{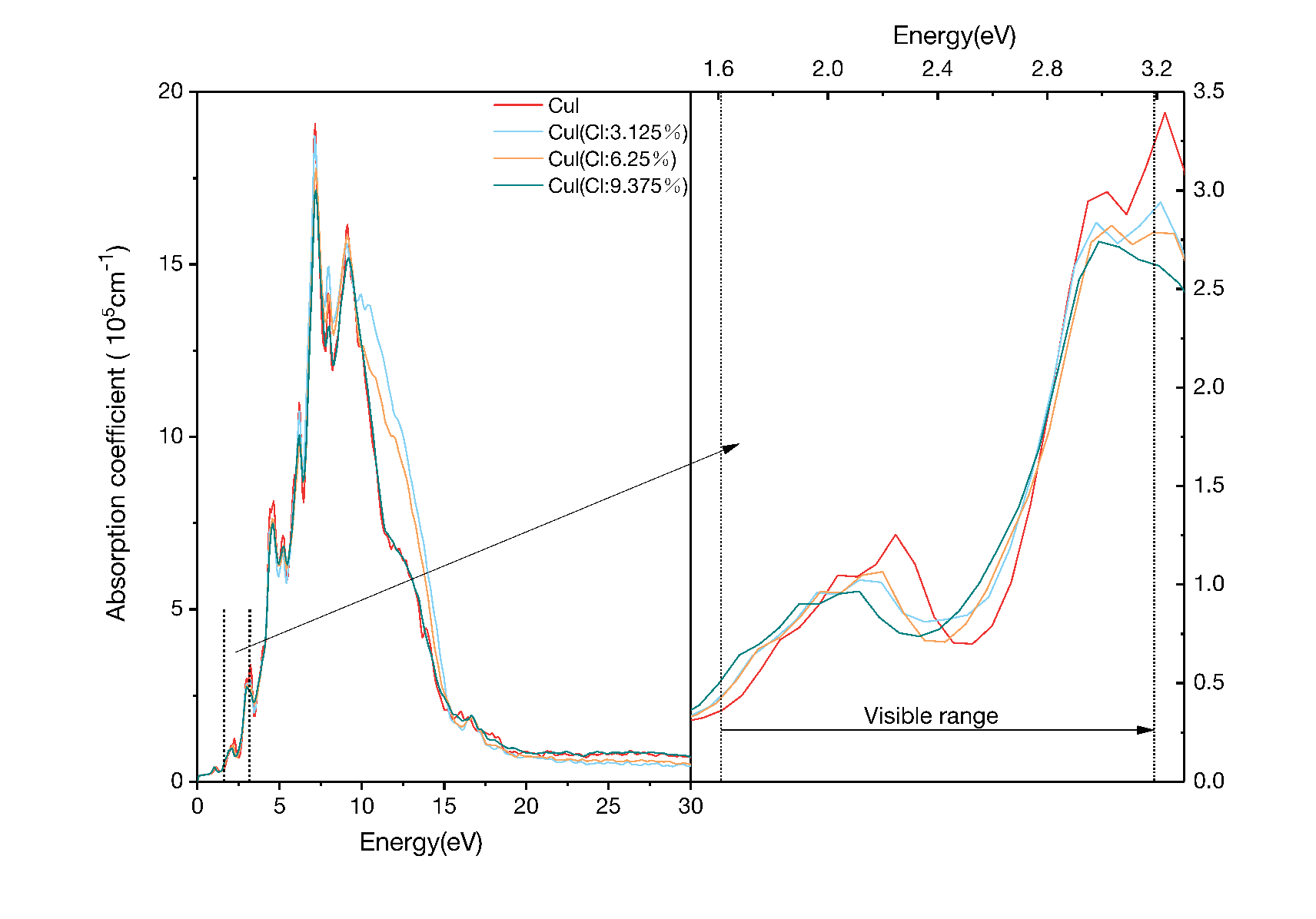}
	  \caption{The absorption coefficient of doping Cl elements}\label{fig18}
\end{figure}

Next, we discuss the light absorption coefficients before and after doping to understand more intuitively the effect of elemental doping on optical properties.

\subsubsection{Absorption coefficient}

\begin{figure}[htpb]
	\centering
		\includegraphics[width=12.0cm]{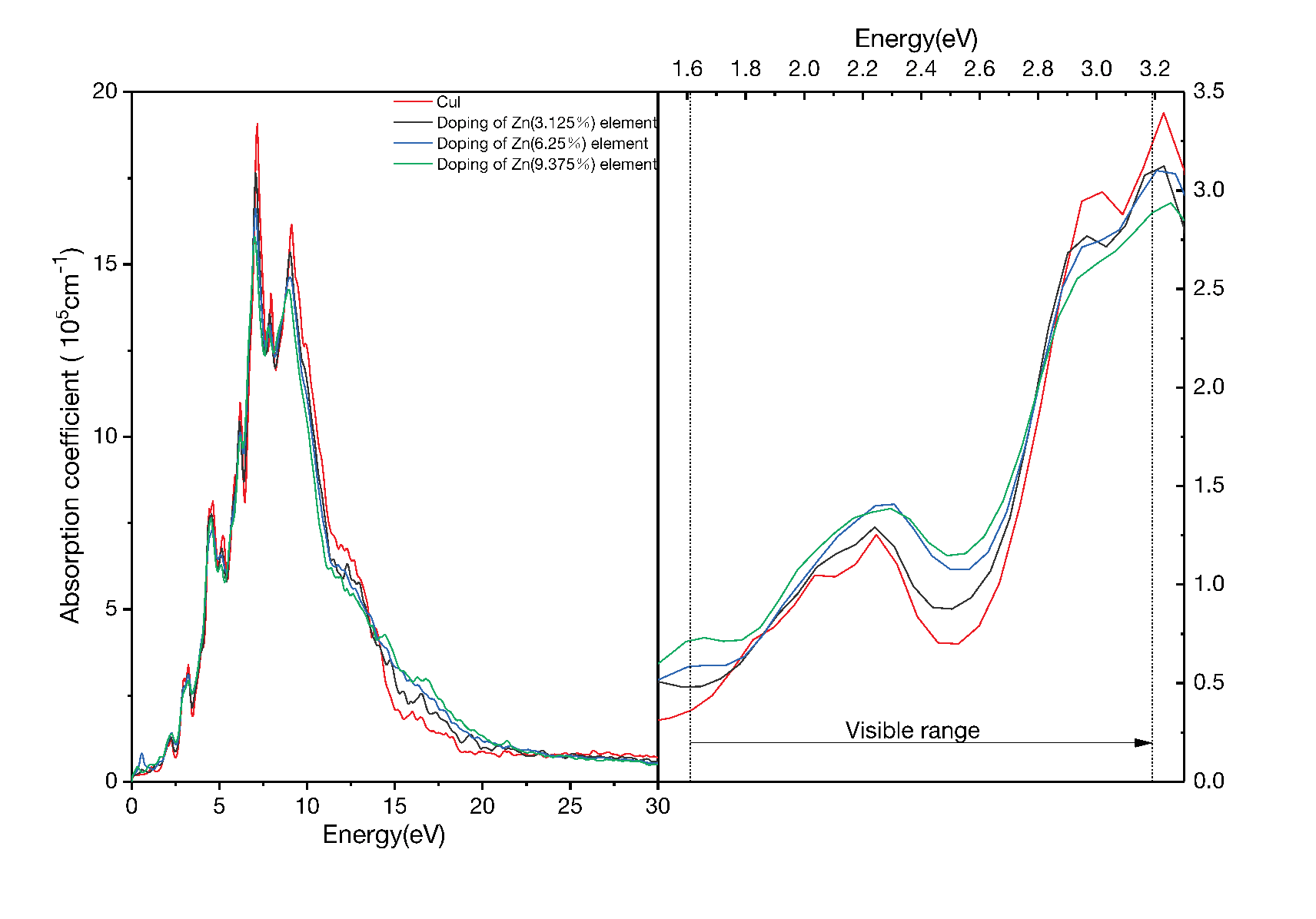}
	  \caption{The absorption coefficient of doping Zn elements}\label{fig19}
\end{figure}

\begin{figure}[htpb]
	\centering
		\includegraphics[width=12.0cm]{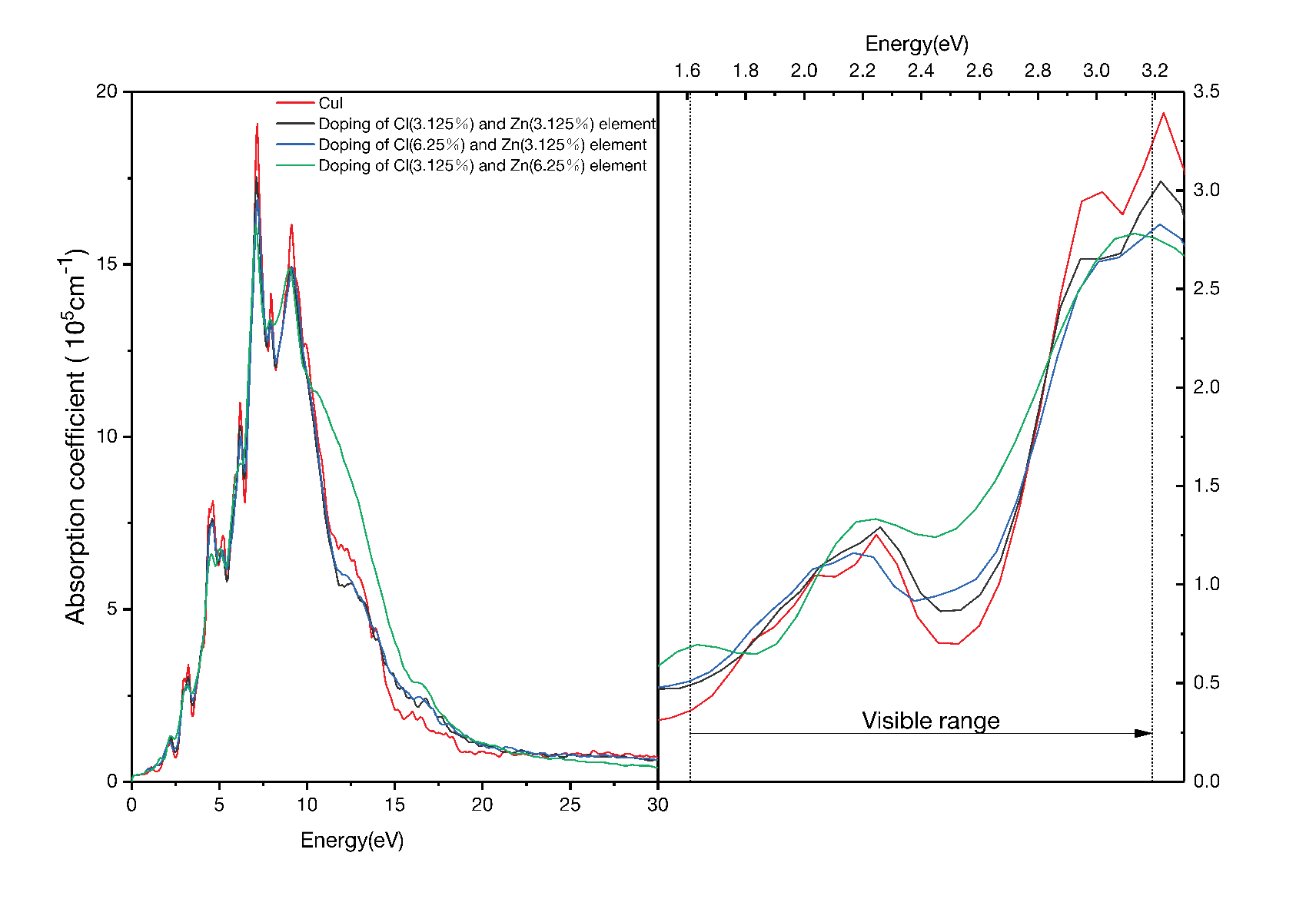}
	  \caption{The absorption coefficient of doping Cl and Zn elements}\label{fig20}
\end{figure}

We can analyze the optical properties of the crystal before and after doping more intuitively from the optical absorption coefficient. Based on the dielectric function, the light absorption coefficient of a $\gamma$-CuI scintillator can be calculated by the following equation.

\begin{equation}\label{5}
  \alpha(\omega)=\sqrt{2\left[-\varepsilon_1(\omega) \pm \sqrt{\varepsilon_1^2(\omega)+\varepsilon_2^2(\omega)}\right]}
\end{equation}

where $\alpha(\omega)$ is the light absorption coefficient, $\varepsilon_1(\omega)$ is the real part of the dielectric function, and $\varepsilon_2(\omega)$ is the imaginary part of the dielectric function. It can be seen that the absorption coefficient is positively correlated with the imaginary part of the dielectric function, and the larger imaginary part of the dielectric function, the stronger the light absorption capacity.

Figure 9 to Figure 11 show the optical absorption coefficients before and after doping, where the total light absorption coefficient is shown on the left and the light absorption coefficient in the visible range is shown on the right. Figure 9 shows the optical absorption coefficients after Cl element doping, it can be seen that in the high energy region, the doping of Cl element makes the absorption coefficient greater than the intrinsic $\gamma$-CuI, while in the low energy region, at the peak of the absorption coefficient, Cl element doping decreases the optical absorption coefficient, and in the visible range, the Cl element doping also makes the light absorption coefficient decrease. The effect of Zn element doping on the light absorption coefficient can be seen in Figure 10, it can be seen that the doping of Zn element is decreasing for light absorption coefficients when the energy is more than 2.8 eV, and the light absorption coefficient is increasing in the visible range and is proportional to the Zn doping concentration, a new absorption peak appears around energy 0.5 eV. Figure 11 shows the absorption coefficients of Cl and Zn elements co-doping. When the Zn concentration is greater than the Cl concentration, the absorption coefficient in the visible range is greater than the intrinsic $\gamma$-CuI. When the Zn concentration is less than or equal to the Cl concentration, the absorption coefficient in the visible range is less than the intrinsic $\gamma$-CuI, and the doping of Cl elements causes the new absorption peak at 0.5 eV due to the 6.25 at\% Zn concentration doping disappears.

The above conclusion is consistent with the conclusion drawn from the imaginary part of the dielectric function.

\subsubsection{Reflection coefficient}

\begin{figure}[htpb]
	\centering
		\includegraphics[width=12.0cm]{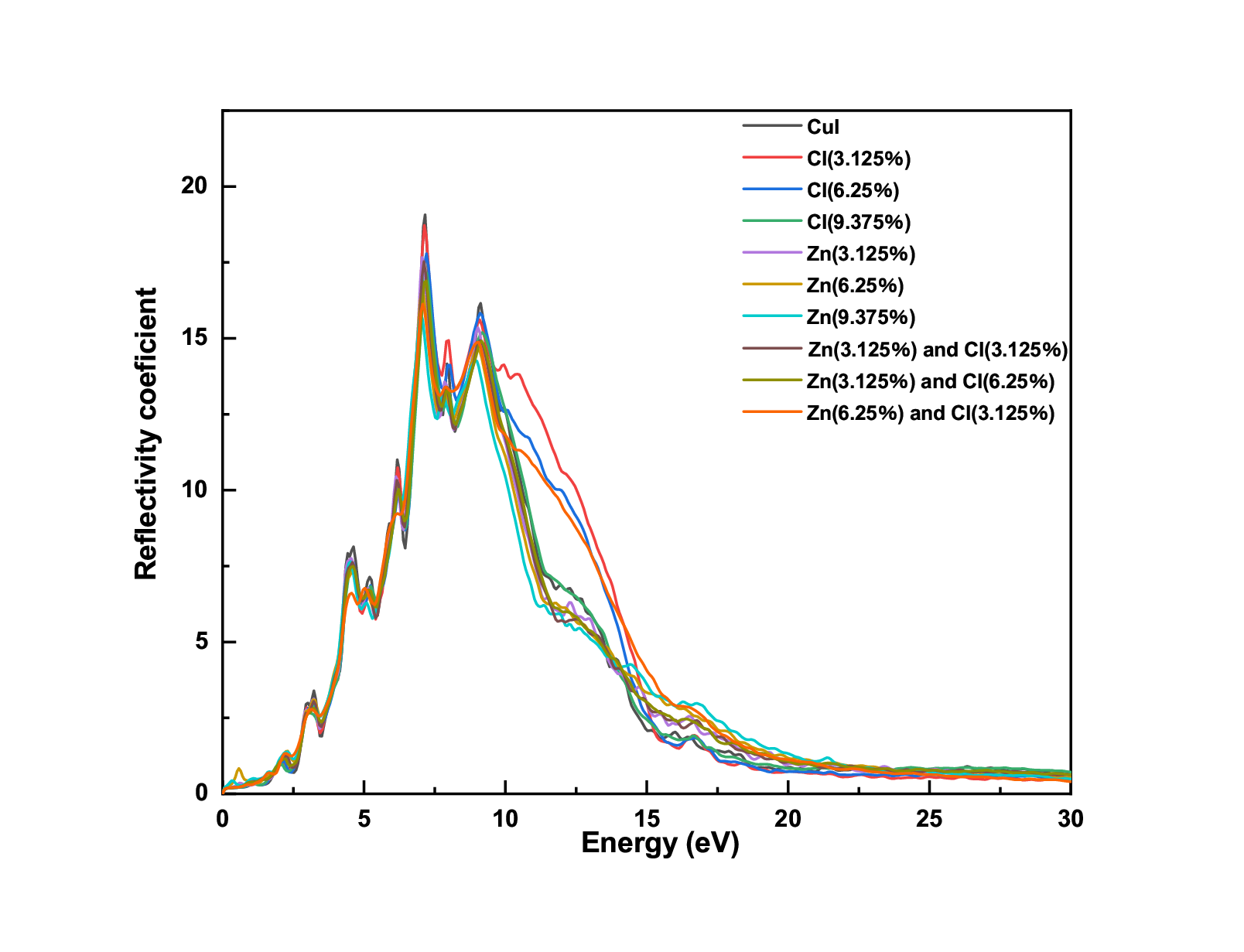}
	  \caption{The reflectivity coefficient of doping elements}\label{fig21}
\end{figure}

\begin{figure}[htpb]
	\centering
		\includegraphics[width=16.0cm]{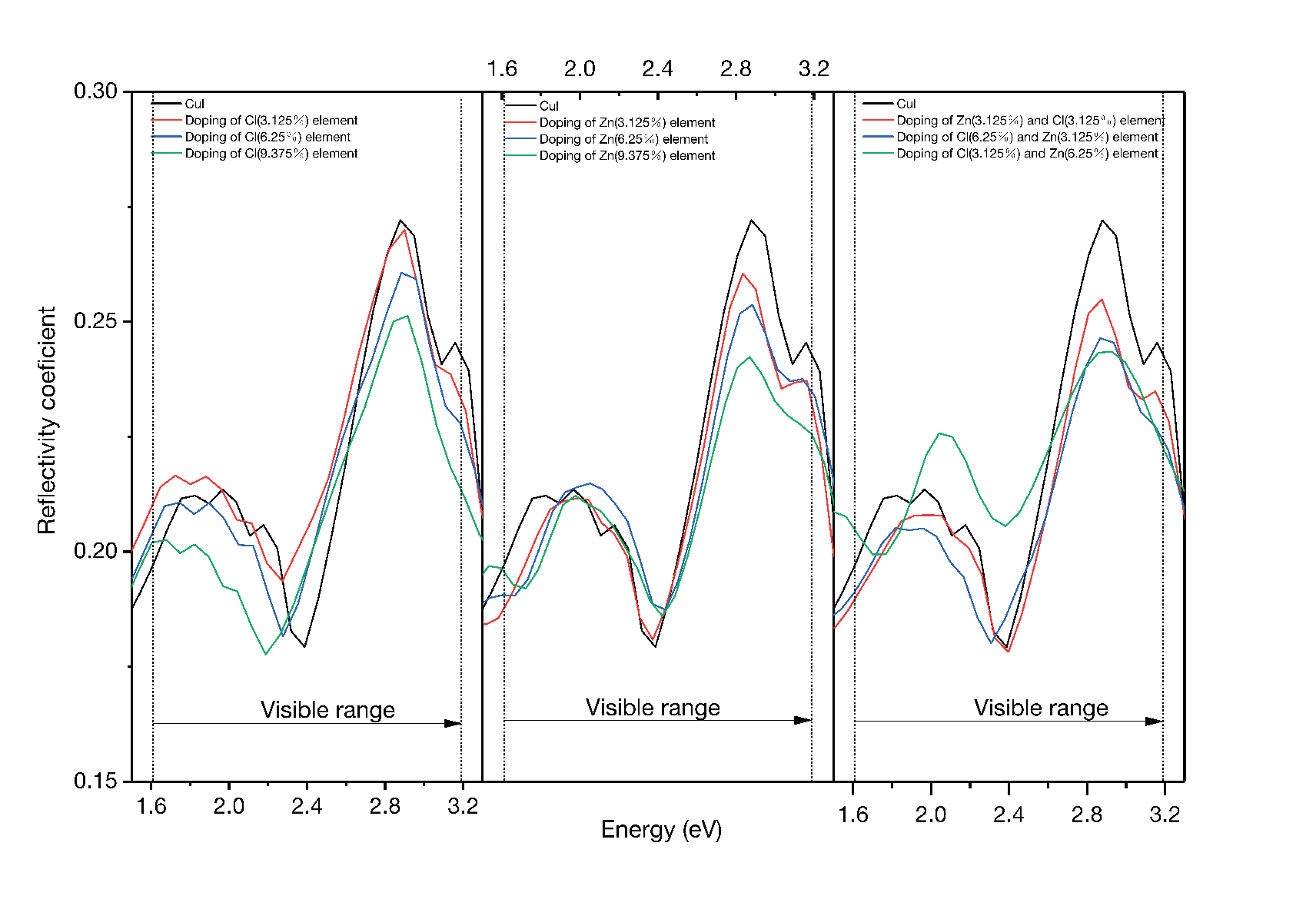}
	  \caption{The reflectivity coefficient of doping elements in visible range}\label{fig22}
\end{figure}

A smaller reflection coefficient allows more photons to be detected by absorption, improving the detection efficiency of $\gamma$-CuI scintillators. Figure 12 shows our calculation of the processing reflection coefficient, and Figure 13 shows the reflection coefficient in the visible range. From Figure 12, it can be seen that 3.125 at\% Cl doping, 6.25 at\% Cl doping and 6.25 at\% Zn and 3.125 at\% Cl co-doping make the reflection peak around 10 eV lower and wider, thus making the reflection coefficient greater when the energy is more than 11 eV and less when the energy is less than 11 eV than the intrinsic $\gamma$-CuI. The reflection coefficients of all other forms of doped crystals are smaller than that of intrinsic $\gamma$-CuI. From Figure 12, it is evident that 3.125 at\% Cl doping, 6.25 at\% Cl doping, and 6.25 at\% Zn and 3.125 at\% Cl co-doping result in a lower and wider reflection peak around 10 eV. As a consequence, the reflection coefficient is greater when the energy exceeds 11 eV and lower when it is less than 11 eV compared to the intrinsic $\gamma$-CuI. The reflection coefficient of other forms of doping is smaller than that of the intrinsic $\gamma$-CuI. In a word, the reflection coefficient of the intrinsic $\gamma$-CuI can be reduced by choosing a suitable co-doping ratio.

\section{Conclusion}

We have calculated the effects of elemental Cl, Zn doping and co-doping on the electronic structure and optical properties of $\gamma$-CuI using first-principles calculations. W find that the doping of Cl elements decreases the band gap and reduces the absorption coefficient in the visible range and is inversely proportional to the doping concentration. This allows more photons to be detected and can improve the scintillation detection performance of $\gamma$-CuI.
Zn element doping will reduce the absorption coefficient when the energy is greater than 2.8 eV, which is inversely proportional to the doping concentration. With the doping concentration increasing, the Fermi energy level from the valence band into the conduction band, 9.375 at\% concentration of doping to show the N-type semiconductor characteristics.
Both Cl and Zn co-doping will reduce the $\gamma$-CuI band gap, and the doping of Cl elements will reduce the red shift caused by Zn doping. When doping $C_{\rm{Cl}}<C_{\rm{Zn}}$, the optical absorption coefficient and the reflection coefficient are smaller than the intrinsic $\gamma$-CuI.

In conclusion, Cl element doping as well as Cl and Zn element co-doping at $C_{\rm{Cl}}<C_{\rm{Zn}}$ can improve the scintillation detection performance of $\gamma$-CuI. The Fermi energy level of 9.375 at\% Zn concentration doping is in the conduction band, which shows N-type semiconductor properties, it provides the basis for making into PN junctions, increasing the application of $\gamma$-CuI in solar and indoor energy harvesting, sensing imaging, etc.

\section*{Declaration of competing interest}
The authors declare that they have no known competing financial interests or personal relationships that could have appeared to influence the work reported in this paper.

\printcredits

\section*{Acknowledgments}
This work is supported by the National Key R\&D Program of China (Grant No. 2022YFB3604804).

%
\bibliographystyle{elsarticle-num}

\bibliography{cas-sc-template.bib}

\end{document}